\documentclass[conference]{IEEEtran}

\usepackage[keeplastbox]{flushend}
\usepackage{multirow}
\usepackage{amsmath,amssymb,amsfonts}
\usepackage{cleveref}
\usepackage[flushleft]{threeparttable}
\usepackage{algorithmic}
\usepackage{graphicx}
\usepackage{textcomp}
\usepackage{booktabs}
\usepackage{graphicx}
\usepackage[caption=false]{subfig}
\usepackage{url}
\usepackage{courier}

\usepackage{lipsum}
\usepackage[table]{xcolor}
\usepackage[style=ieee, doi=false]{biblatex}

\newcommand{\action}[1]{\texttt{#1}} 

\usepackage[top=0.75in, bottom=1.041in, left=0.68in, right=0.646in]{geometry}

\def\BibTeX{{\rm B\kern-.05em{\sc i\kern-.025em b}\kern-.08em
    T\kern-.1667em\lower.7ex\hbox{E}\kern-.125emX}}

\usepackage{pgfplots}
\pgfplotsset{compat=1.5}
\usetikzlibrary{intersections} 
\usepgfplotslibrary{external} 
\begin{document}


\title{Prioritizing emergency evacuations under \\ compounding levels of uncertainty\\}

\author{\IEEEauthorblockN{Lisa J. Einstein$^1$, Robert J. Moss$^1$, Mykel J. Kochenderfer$^{1,2}$}
\IEEEauthorblockA{\textit{$^{(1)}$Department of Computer Science, $^{(2)}$Department of Aeronautics and Astronautics} \\
Stanford University, Stanford, California, USA \\
\{lisae, mossr, mykel\}@cs.stanford.edu}
}

\maketitle

\begin{abstract}
 Well-executed emergency evacuations can save lives and reduce suffering.
 However, decision makers struggle to determine optimal evacuation policies given the chaos, uncertainty, and value judgments inherent in emergency evacuations. 
 We propose and analyze a decision support tool for pre-crisis training exercises for teams preparing for civilian evacuations and explore the tool in the case of the 2021 U.S.-led evacuation from Afghanistan. 
 We use different classes of Markov decision processes (MDPs) to capture compounding levels of uncertainty in
 (1) the priority category of who appears next at the gate for evacuation,
 (2) the distribution of priority categories at the population level, and
 (3) individuals' claimed priority category.
 We compare the number of people evacuated by priority status 
 under eight heuristic policies. 
 The optimized MDP policy achieves the best performance compared to all heuristic baselines. We also show that accounting for the compounding levels of model uncertainty incurs added complexity without improvement in policy performance.
 Useful heuristics can be extracted from the optimized policies to inform human decision makers.
 We open-source all tools to encourage robust dialogue about the trade-offs, limitations, and potential of integrating algorithms into high-stakes humanitarian decision-making. 
\end{abstract} 


\section{Introduction}
Evacuating people in emergency situations ranging from wars to natural disasters is a complex challenge that entails optimizing for multiple competing objectives under immense uncertainty \cite{8569957, keneally2016markov}. Doing so effectively can save lives and reduce suffering. 
Evacuation problems could be aided by decision-making algorithms that can account for dynamic sources of uncertainty while balancing multiple objectives, a task that can be very challenging for an individual human or even an institution to manage.

We can frame evacuation problems as maximizing some reward, such as saving as many people as possible or rescuing those most likely to be impacted adversely by a crisis. Many evacuation problems also deal with the sensitive issue of prioritizing people when resources are finite. For example, when dispatching helicopters for medical evacuations in a war zone, military officials 
can prioritize requests based on their urgency and the projected number of casualties instead of on a first-come first-served basis \cite{rettke2016approximate}. 

Markov decision processes (MDP) can generate policies that assist human decision makers in making optimal sequential decisions while accounting for limited time and resources. \cite{kochenderfer2015decision, keneally2016markov}. Partially observable MDPs (POMDPs), MDPs where the true state is uncertain, can also be useful for modeling dynamic resource allocation problems such as wildfire management and hospital admittance \cite{diao2020uncertainty, bertsimas2017mcts, griffith2017automated, lee2021multi}. 

We explore the use of different classes of MDPs to guide optimal decision making in the specific context of the U.S.-led evacuation from Afghanistan. As it became clear that the Taliban would seize Afghanistan's capital in August 2021, the White House tasked the Departments of Defense and State with evacuating ``U.S. Embassy personnel, U.S. citizens, and allied personnel with whom the U.S. Government has agreements for evacuation" 
from Afghanistan at a rate of 5000 per day \cite{NSC}. This included Afghans eligible for Special Immigrant Visas (SIVs) due to their contributions to the U.S. effort in Afghanistan. These decisions became increasingly complex over time. When the Taliban came into power, tens of thousands of Afghans were at risk of persecution due to their affiliations or genders and thus also sought evacuation \cite{UNHCR}. The United States ultimately evacuated nearly 124,000 people, making it the largest noncombatant evacuation operation (NEO) in history \cite{Blinken_HFAC, army_article}.
The large crowds of people and time pressure at the airport made it challenging to decide who should be let in and put on aircraft \cite{apocalyptic}. Marines protecting the airport entrances and consular officers responsible for screening the evacuees had to quickly decide who to let through gates and who to turn away based on instructions from the U.S. government and also their own intuition \cite{HART}. 

We model the evacuation problem using different classes of MDPs to capture compounding levels of uncertainty and solve for optimal or approximately optimal policies in the case of the U.S. evacuation from Afghanistan.  We compare the number of people evacuated by priority status and aggregate cumulative reward that was generated by our algorithm with eight heuristic policies.
We demonstrate that an optimized MDP policy outperforms all heuristic policies (simple rules, such as evacuate all American citizens first).
Results also indicate that accounting for the compounding levels of uncertainty in the population distribution and uncertainty in the individual claimed priority status provide little to no added benefit compared to the optimized MDP policy that accounts only for the uncertainty in who will arrive at the gate given a known population distribution.
We open-source all tools to encourage robust dialogue about the trade-offs, limitations, and potential of integrating algorithms into high-stakes decision-making. We target the 2021 Afghanistan evacuation as a demonstration of these techniques, but our contributions could be generalized to other emergency evacuation efforts.

\section{Methodology} \label{methodology}
Using a principled mathematical framework for sequential decision making problems, we can formulate and solve the problem of optimally deciding who to evacuate from a chaotic central location with access to air transportation but limited aircraft seats as was the case of the 2021 Afghanistan evacuation.
We formulate the emergency evacuation problem using different classes of Markov decision processes with varying levels of uncertainty and generate optimal MDP policies using value iteration \cite{kochenderfer2015decision} and approximately optimal POMDP policies using the POMCP solver \cite{pomcp}.

During the evacuation from Kabul, thousands of people crowded the gates at Hamid Karzai International Airport (HKIA), from which the U.S government based its efforts to evacuate people from Afghanistan. Marines guarded the gates and conducted crowd control, selectively permitting small groups to enter. Further inside, consular officers conducted more thorough vetting and decided who could proceed to the airport terminal. The decision-making for the Marines and consular officers was based on centralized policies, but both had some leeway in their decisions. 
As we modeled the interaction, the Marines decided whether to accept or reject each person based on their priority category, the size of their family, remaining capacity in the aircraft, and the time left for the aircraft to depart.
We assume that not all families who enter the airport will board the aircraft.

\subsection{Sequential Problem Formulation}
MDPs and POMDPs consist of a state space, action space, transition function, and reward function. POMDPs also have an observation model that specifies the probability of making a particular observation given the true state (see \cref{fig:claims}).
A discount factor $\gamma$ is used to control how myopic the decisions will be, and in this problem we set $\gamma=1$ so our current decision is dominated by potential future rewards.

\textbf{\textit{States:}}\label{B}
We consider a discrete state space $\mathcal{S}$ where an individual state $s \in \mathcal{S}$. The state $s$ is defined by the following:
\begin{itemize}
  \item The $c$ remaining seats on the aircraft (maximum of $500$).
  \item The $t$ time steps remaining before the flight departs (total time of $1200$ steps).
  \item The family size $f$ of the family aiming to enter the airport to board the aircraft (ranging from $1$ to $13$).
  \item The priority category $v$ of the family wanting to enter the airport ($5$ total categories).
  The categories range in value depending on the family's status priority to the U.S. government, shown in \cref{table:population}. 

\end{itemize}
The combination of these values represents a single state $s$ as a vector $[c,t,f,v]$ with a state space size of $|\mathcal{S}|=40{,}047{,}346$.

The initial state of the MDP has $t$ and $c$ initialized to the total time remaining and total available seats on the aircraft, respectively. We randomly generated the family size $f$ and priority status $v$ of the first family and make the assumption that all family members share the same priority category. 

While population estimates were challenging to calculate during the evacuation, we used values estimated by the Association of Wartime Allies. See \cref{table:population} for population estimates.
\begin{table}[!ht]
    \centering
    \caption{\label{table:population} Population estimates and associated rewards for attempted arrivals to the airport by claimed status.}
    \begin{threeparttable}
    \begin{tabular}{lrr}
    \toprule
    \textbf{Priority Category} & \textbf{Est. Population}\tnote{*} & \textbf{Reward}\\
    \midrule
    AMCIT\tnote{1} & $14786$  & $100$  \\ 
    \midrule
    SIV\tnote{2} & $123000$ &  $25$ \\ 
    \midrule
    P1/P2 Afghan\tnote{3} & $604500$ & $5$ \\ 
    \midrule
    Vulnerable Afghan\tnote{4} & $1000000$ & $1$ \\
    \midrule
    ISIS-K\tnote{5} & $20$ & $-500$\\
    \bottomrule
    \end{tabular}
    \begin{tablenotes}
        \item[1] {AMCITS stands for American Citizens.}
        \item[2] {SIV holders performed activities with and for U.S. military personnel.}
        \item[3] {P1/P2 stands for Priority 1 and 2, and includes Afghans and their immediate family members who may have been at risk due to their U.S. affiliation but who are not eligible for a Special Immigrant Visa (SIV) \cite{State}.}
        \item[4] {Afghans without claims for priority humanitarian parole but who are otherwise vulnerable due to affiliations.}
        \item[5] {ISIS-K is a terrorist organization based in Afghanistan.}
        \item[*] {The method used to estimate the population is described in supplementary documentation of the code: \url{https://github.com/sisl/EvacuationPOMDP.jl}}
    \end{tablenotes}
    \end{threeparttable}
\end{table}



\textbf{\textit{Actions:}}
Our discrete action space $\mathcal{A}$ accounts for two possible actions: if the family is \action{ACCEPTED} or \action{REJECTED} by the Marine who they approach at the airport gate (which does not necessarily mean they board the aircraft if they are accepted into the airport).

\textbf{\textit{Transitions:}}
To deal with the uncertainty of transitioning from the airport to the aircraft, we define the transition function, denoted $T(s' \mid s, a)$, to determine the probability of transitioning to some future state $s'$ when taking action $a$ from state $s$.
The aircraft capacity and the time variables each start at their maximum number of seats available and total time, respectively.
When either reaches zero the problem terminates. If the family being considered has a size exceeding the current capacity of the aircraft, we accept up to a single family over capacity and then terminate. Each future family has a probability of being a certain size $f$ and priority $v$. If the Marine chooses to \action{ACCEPT} a family into the airport, we assume there is an $80\%$ chance that the family ultimately boards the flight. 
However, if the action is \action{REJECT}, the family will not make it onto the flight.

\textbf{\textit{Rewards:}}
A reward is associated to each state-action pair, which is denoted $R(s,a)$.
When the action is to \action{REJECT}, we give a reward of zero.
Otherwise, when the action is to \action{ACCEPT}, the total reward is $f v + \epsilon$, the family size times the associated priority status reward (see \cref{table:population}) plus a small value of $\epsilon=10^{-4}$ to settle any ties on the decision boundaries in favor of \action{ACCEPT}. As shown in \cref{table:population}, we define rewards associated with each priority status based on U.S. priorities expressed in statements and clarified with government officials. While we recognize the sensitive nature of ranking humans in order of priority, these were the parameters that decision makers at the White House had to consider given their limited time and capacity. We do not assume our model parameters are strictly correct. Our model of the problem allows for this rank order and associated rewards to be easily changed based on expert information.

\subsection{Compounding Uncertainty}

\begin{figure}
    \centering
    \includegraphics[width=\columnwidth]{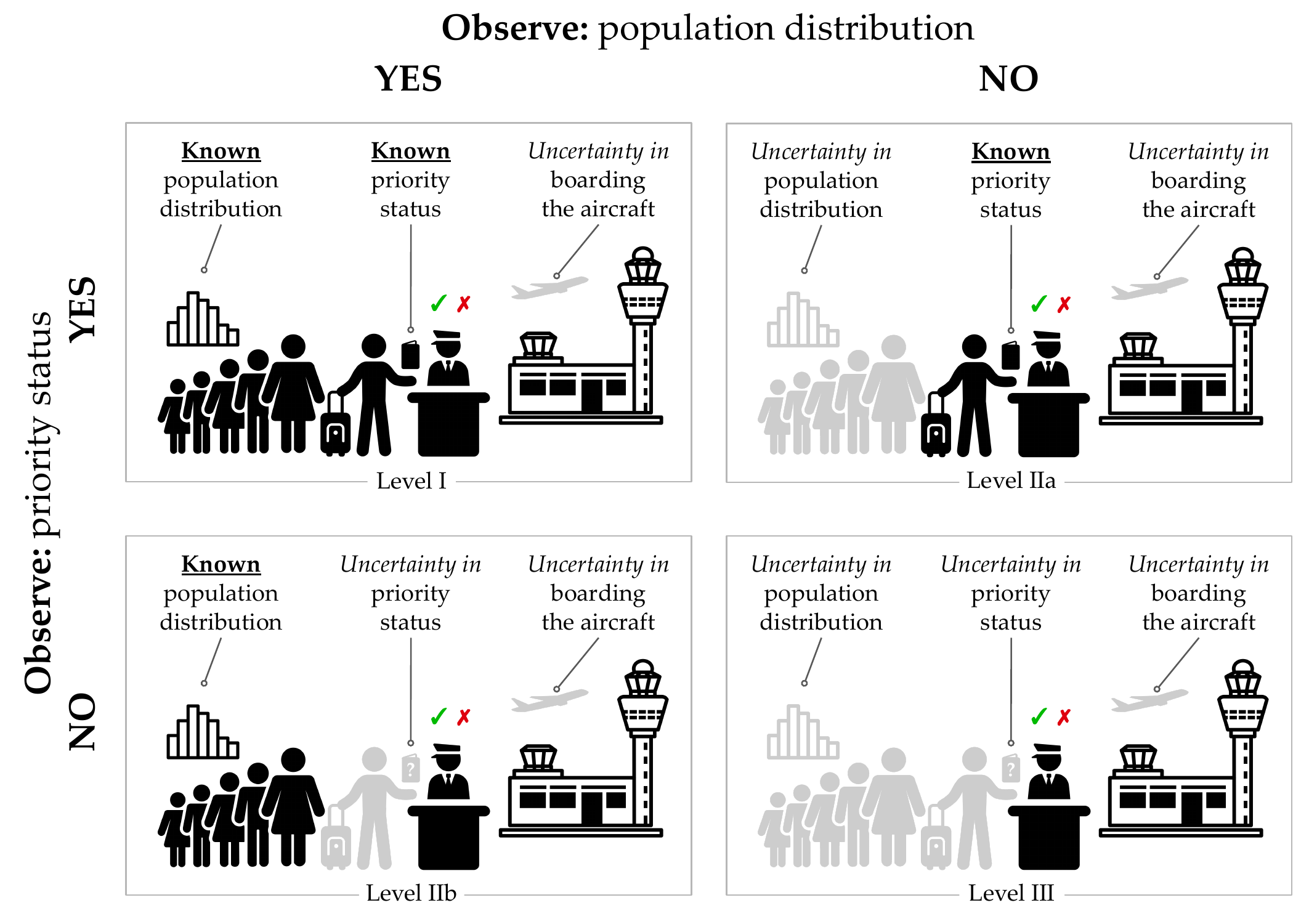}
    \caption{Compounding levels of uncertainty using different classes of MDPs. All levels incorporate uncertainty about who will arrive next at the gate.}
    \label{fig:uncertainty_matrix}
\end{figure}

\begin{figure*}
    \centering
    \setkeys{Gin}{width=0.34\textwidth}
    \subfloat[Family size distribution at the gate.\label{fig:family_size}]
        {\includegraphics{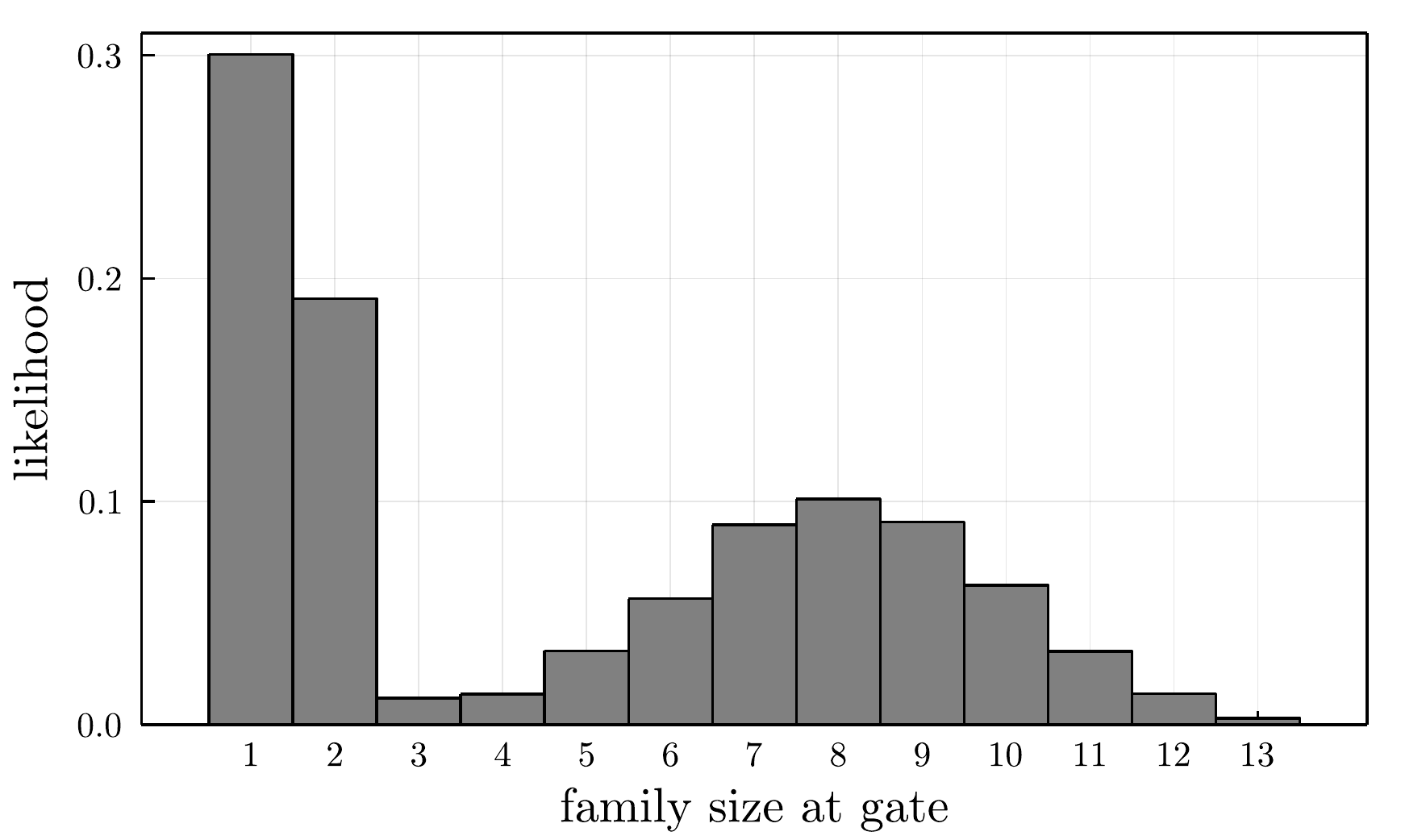}}
    \subfloat[Population distribution of priority statues.\label{fig:pop_distr}]
        {\includegraphics{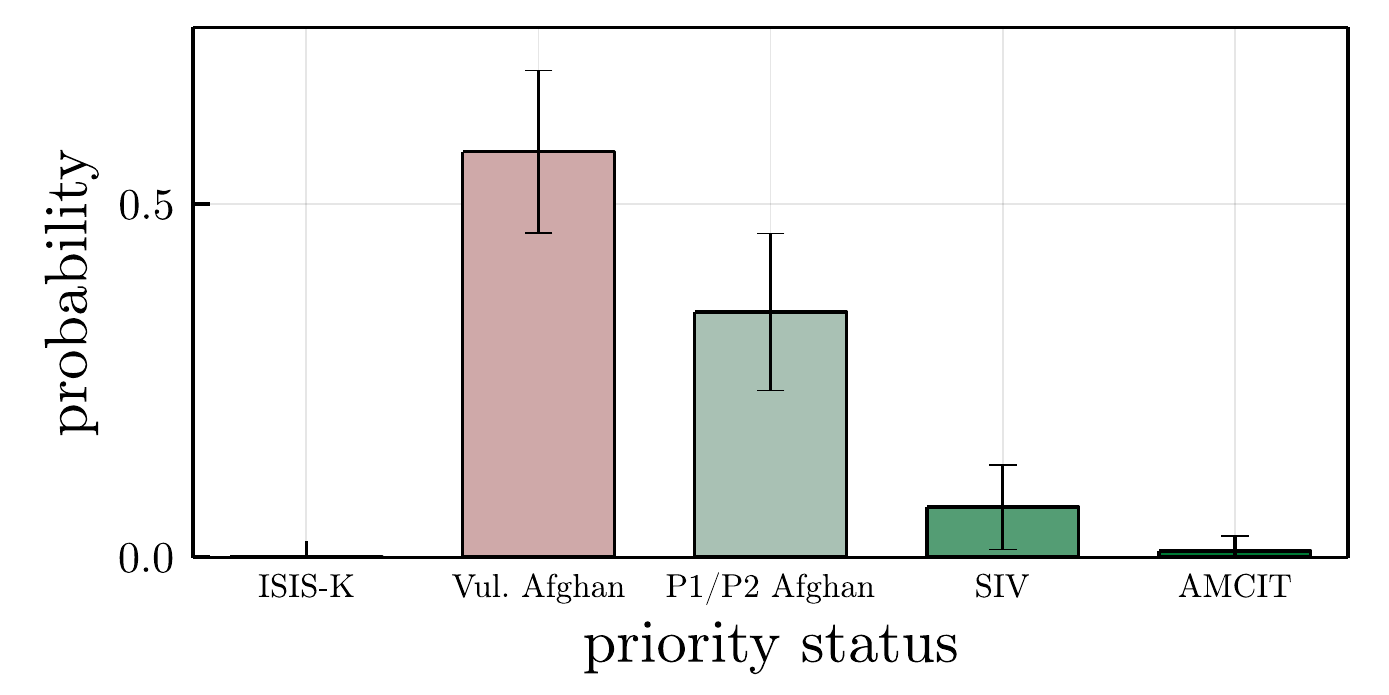}}
    \subfloat[Claimed priority status models.\label{fig:claims}]
        {\resizebox{0.25\linewidth}{!}{\includegraphics{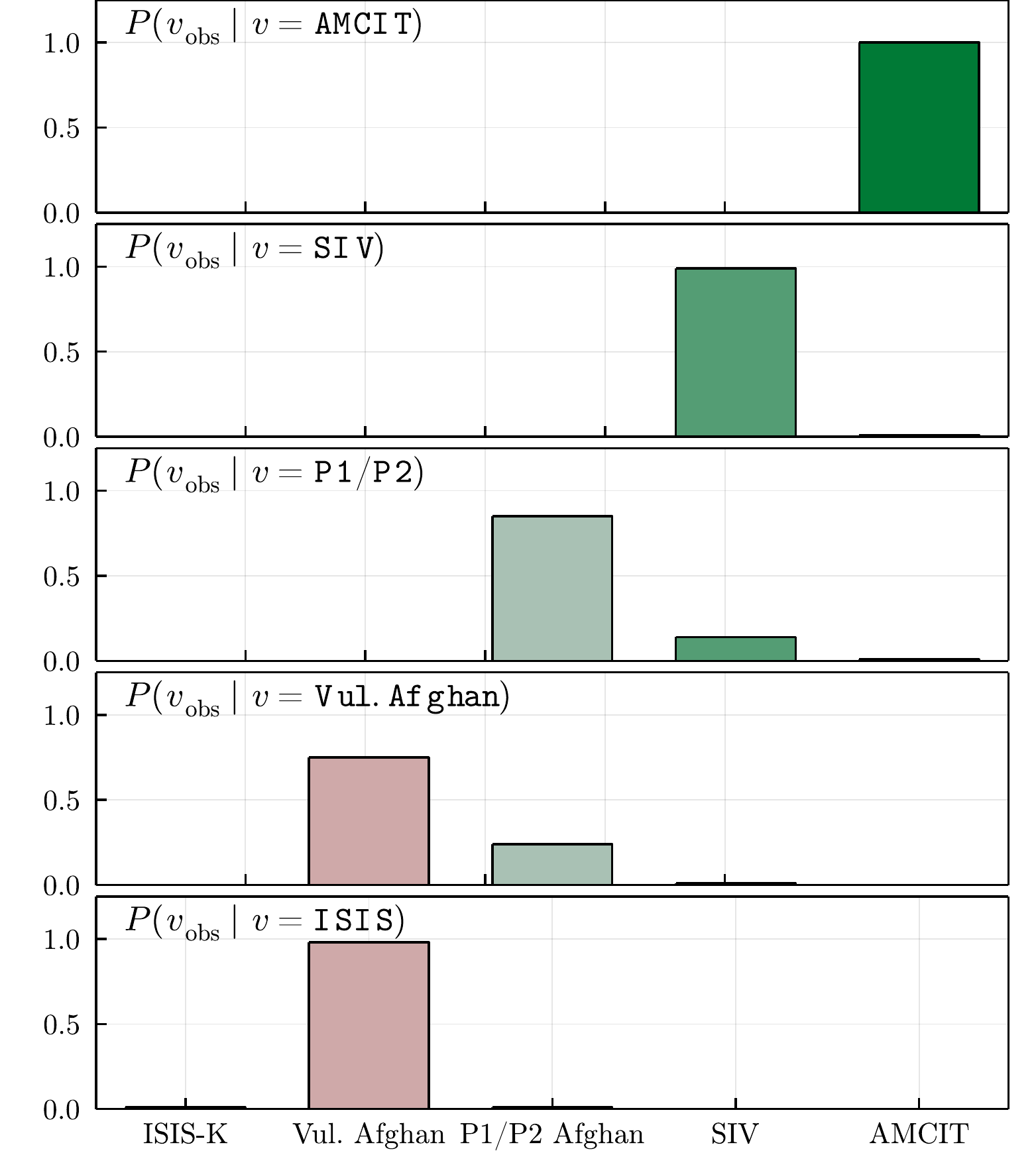}}}
    \caption{Models used in the problem formulation.}
    \label{fig:models}
\end{figure*}

We investigate optimal evacuation strategies using four compounding levels of uncertainty as seen in \cref{fig:uncertainty_matrix}. 
Subsequent levels include the uncertainties from the previous levels.

\textbf{Level I:}
The problem is modeled as an MDP where we assume full observability of the population distribution and assume that the claimed status at the gate is their true priority status. 
Uncertainty is used in the transition function that includes (1) the priority status of who we expect to arrive at the gate next and (2) whether the family accepted into the airport will board the aircraft.

\textbf{Level IIa:} We make all the same assumptions as level I, but add uncertainty about the true population distribution. We model level IIa as a POMDP (technically, a POMDP-lite with the population distribution as the hidden parameter \cite{pomdp_lite}).
We have an estimate of the population distribution in the beginning (shown in \cref{fig:pop_distr}) and then update it as we see people arrive at the gate.
To update the population distribution over time, we use a Dirichlet distribution as our belief over the true priority statuses. By updating this distribution, we can converge to the true distribution by the end of the simulation.

\textbf{Level IIb:} We assume observability in the population distribution, but not in the observed state of the priority status.
In reality, priority status is never known with total certainty.
As an example, Marines and consular officers had to make priority status judgements based on a wide range of documents provided by the families, from passports to crumpled papers to emails on phones to covert symbols.
Our \textit{claim model} incorporates the likelihood that someone saying they are in a particular priority category actually belongs to that claimed priority category (shown in \cref{fig:claims}).
Given the ranked nature of ``priorities'', a particular claimed status of a family has a majority of the probability mass associated with their true status, while the remainder of the probability mass is distributed to the right (i.e., claiming higher priority than the truth, thus increasing the likelihood of getting into the airport).
This is true for all priorities except those belonging to ISIS-K, as it is unlikely that members of a terrorist organization would disclose their true category. 

We could model level IIb as a POMDP because we have state uncertainty over the true priority status, but
since the belief over the priority status is not updated based on subsequent observations, we can simplify the problem and model it as an MDP with an observation-based weighted reward function:
\[R(o,a) = \sum_{s\in\mathcal{S}} P(s \mid o) R(s,a)\]
where $P(s \mid o)$ is the likelihood of the true state $s$ given the observation $o$, computed from Bayes' rule using the probability distributions in \cref{fig:claims}.
Formulating this as an MDP means we can solve the problem exactly for an optimal policy.

\textbf{Level III: } We assume uncertainty in both the population distribution and claimed priority status. We combine uncertainties described in levels I, IIa, and IIb and model the problem as a POMDP for the same reasons as level IIa (i.e., a requirement on updating the population belief over time).


\subsection{Assumptions} We made a number of assumptions in order to model the evacuation, all of which can be fine-tuned with increased expert knowledge. 
 
\textbf{Family size.} We modeled family size as a multi-modal distribution using a mixture of two truncated Gaussian distributions shown in \cref{fig:family_size}.
The first Gaussian distribution $\mathcal{N}(8,2)$ has a mean and standard deviation that reflects the average family size in Afghanistan \cite{prb}.
The second Gaussian distribution $\mathcal{N}(1, 0.6)$ reflects individuals or couples who might arrive at the gate without their entire family.

\textbf{Estimated population distribution at airport.} We estimated the number of people at the airport by priority category using numbers calculated by the Association of Wartime Allies \cite{AfghanAllies_data} (see \cref{fig:pop_distr}). We included these people as possible arrivals at the gate, though some percentage likely decided to stay home for safety reasons.

\textbf{Distribution of ISIS-K, a terrorist organization based in Afghanistan.} We assumed that the likelihood that ISIS-K would arrive at the gate was a low-probability, high-consequence event. We assumed about $20$ ISIS-K members circulated the gates attempting to enter the airport.

\textbf{Priority categories of families.} For this exercise, we assumed that immediate families shared the same priority category and that a person accepted would be able to bring all family members present with them. During the evacuation, U.S. officials did try to keep immediate family together to the degree possible, even if dependents of primary visa holders did not individually hold the same priority status.


\textbf{Rewards stay constant.} Rewards for varying priority remain constant throughout the exercise. In reality, as the security situation developed, the rewards associated with each priority status adapted as well.
Problems that include a reward function that changes over time can be addressed using non-stationary MDPs \cite{lecarpentier2019non}.

\textbf{Time remains constant.} We assumed that the time horizon was constant and known. Given the quickly evolving situation during the evacuation, there was not always certainty about the amount of time U.S. forces would have to evacuate people. 



\subsection{Solution Methods}
In the context of MDPs, a policy maps states (or observations) to actions, thus providing a strategy to look up the recommended action to take given the current state of the problem.
Given that the levels differ in their problem formulations (i.e., choice of MDP class), we can use different solvers to optimize policies for each level.

Levels I and IIb, defined as MDPs, can be solved exactly using value iteration to obtain an optimal policy as an offline look-up table \cite{kochenderfer2015decision}. Level IIb can be solved exactly because our belief about our claim is not changing over time. We account for the claims model by updating the reward function to be a weighted average of the likelihood a family falls into a particular status multiplied by the reward they would receive for each status.

In both level IIa and level III, we update our population distribution over time. To optimize our choice of action, we use an online POMDP solver called POMCP (partially observable Monte-Carlo planning) \cite{pomcp}. We use a max depth of $120$ and $500$ iterations, which we chose to balance runtime and policy performance.
When optimizing the policies using POMCP, instead of using a random or heuristic rollout policy to estimate the value of a particular state-node in the search tree, we use the previously optimized MDP policies to provide the exact value of a given state, thus greatly reducing the variance in the state-value estimate.

\section{Experiments} \label{experiments}

We generate 1000 static state-observation trajectories to provide each policy with comparable settings under which to be evaluated. Static trajectories include the state observed by the Marine and the actual state of the person. We generate the 1000 trajectories by sampling from the Dirichlet distribution in \cref{fig:pop_distr}, so the population distribution we sample from in simulation is different each time. We run 1000 different simulations given the same observed state and then aggregate the cumulative reward of that policy used on the 1000 simulations.
Each policy is shown the \textit{observed} state and never the true state so we can compare the performance in a realistic setting (where the Marines would never know the true state exactly, but would only be able to make imperfect observations).

We solve for optimized policies integrating each level of uncertainty.
The choice of solver has runtime and performance impacts that may contribute to the choice of the MDP class when modeling similar problems.
We compared the optimized policies against eight baseline heuristic policies:
\begin{enumerate}
    \item  \textbf{After Threshold AMCITs}: Make decisions based on our optimized MDP policy until there are 200 time steps remaining. Then, accept American citizens only. 
    \item \textbf{Before Threshold AMCITs}: Accept only American citizens until there are 200 time steps left and then make decisions based on the optimized MDP policy.
    \item \textbf{AMCITS}: Only accept American citizens.
    \item \textbf{SIV and AMCITs}: Accept only American citizens or Special Immigrant Visa (SIV) applicants.
    \item \textbf{SIV, AMCITs, and P1P2s}: Accept only American citizens, Special Immigrant Visa (SIV), or Priority 1/2 applicants.
    \item \textbf{Non-ISIS-K}: Accept everyone except ISIS-K members. Note that given the claims models, there is a very low probability that an actual ISIS-K member will claim their true category.
    \item \textbf{Accept All}: Accept anyone who comes to the door.
    \item \textbf{Random}: Randomly accept or reject whoever arrives at the door with equal probability.
    
\end{enumerate}
We built upon code from the POMDPs.jl framework \cite{pomdps_jl} written in the Julia programming language and have open-sourced all of the code including the evacuation modeling tools, simulators, and policies.\footnote{\url{https://github.com/sisl/EvacuationPOMDP.jl}}

\section{Results}
\setlength{\tabcolsep}{5pt}

\begin{table*}[!ht]
    \centering
    \scriptsize
    \caption{Experiment Results}
    \label{tab:results}
    \begin{threeparttable}
    \begin{tabular}{@{}lrrrrrrr@{}}
    \toprule
    \textbf{Policy} & \textbf{Reward} & \textbf{Airport Accepted/Total\tnote{$\dagger$}} & \textbf{AMCIT\tnote{*}} & \textbf{SIV\tnote{*}} & \textbf{P1/P2\tnote{*}} & \textbf{Afghan\tnote{*}} & \textbf{ISIS-K\tnote{*}}  \\
    \midrule
    Level I                         &               $12112.97 \pm 253.96$   &               $(629.50 \pm 1.06)/5438.38$             &       $41.79/42.35$   &       $260.25/377.87$ &       $275.05/1895.42$        &       $52.41/3122.46$         &       $0/0.28$                \\
    Level IIa (approx.)             &               $11003.86 \pm 233.69$   &               $(630.07 \pm 0.98)/5348.84$             &       $40.13/40.38$   &       $220.03/369.80$ &       $280.20/1864.41$        &       $89.71/3073.97$         &       $0/0.28$                \\
    Level IIb                       &               $12120.23 \pm 254.28$   &               $(629.57 \pm 1.06)/5445.13$             &       $41.88/42.43$   &       $260.23/378.50$ &       $274.89/1897.67$        &       $52.58/3126.25$         &       $0/0.28$                \\
    Level III (approx.)             &               $9664.10 \pm 250.84$    &               $(630.17 \pm 1.02)/5505.53$             &       $40.26/43.6$    &       $161.85/384.60$ &       $291.66/1920.12$        &       $136.38/3156.92$        &       $0.01/0.28$             \\
    \midrule
    AfterThresholdAMCITs            &               $11401.82 \pm 263.34$   &               $(542.56 \pm 2.55)/5524.68$             &       $42.20/42.68$   &       $240.58/387.04$ &       $226.92/1926.68$        &       $32.86/3167.98$         &       $0/0.29$                \\
    BeforeThresholdAMCITs           &               $8495.33 \pm 347.00$    &               $(622.02 \pm 1.04)/5638.80$             &       $50.66/50.66$   &       $67.33/395.93$  &       $311.89/1964.45$        &       $192.12/3227.47$        &       $0.01/0.29$             \\
    AMCITs                          &               $5308.43 \pm 365.61$    &               $(77.80 \pm 3.67)/5653.04$              &       $51.09/51.09$   &       $3.94/397.38$   &       $19.53/1970.23$         &       $3.24/3234.06$          &       $0/0.29$                \\
    SIV-AMCITs                      &               $11400.89 \pm 230.53$   &               $(566.86 \pm 3.23)/4598.76$             &       $35.57/35.57$   &       $264.75/264.75$ &       $239.60/1602.52$        &       $26.93/2695.75$         &       $0/0.17$                \\
    SIV-AMCITs-P1P2                 &               $4906.96 \pm 82.13$     &               $(627.93 \pm 1.11)/1125.44$             &       $9.78/9.78$     &       $75.39/75.39$   &       $375.28/375.28$         &       $167.48/664.94$         &       $0/0.06$                \\
    Non-ISIS-K                        &               $3096.94 \pm 55.39$     &               $(627.70 \pm 1.09)/627.70$              &       $5.60/5.60$     &       $43.97/43.97$   &       $219.45/219.45$         &       $358.65/358.65$         &       $0.04/0.04$             \\
    AcceptAll                       &               $3096.94 \pm 55.39$     &               $(627.70 \pm 1.09)/627.70$              &       $5.60/5.60$     &       $43.97/43.97$   &       $219.45/219.45$         &       $358.65/358.65$         &       $0.04/0.04$             \\
    Random                          &               $3093.61 \pm 53.48$     &               $(627.42 \pm 1.07)/1257.04$             &       $5.44/11.59$    &       $44.29/87.50$   &       $219.34/439.67$         &       $358.32/718.22$         &       $0.02/0.06$             \\
    \bottomrule
    \end{tabular}
    \begin{tablenotes}
        \item[$\dagger$] {Total number of people accepted into the airport which could be larger than the capacity of the aircraft (500) due to the transition probability of $0.8$.}
        \item[*] {Claimed status presented to each policy while the true status is recorded in the metrics.}
    \end{tablenotes}
    \end{threeparttable}
\end{table*}

The MDP formulation is optimal as it accounts for a number of different factors.
The heuristic policies use a rule-based approach to decide who to let in rather than operating as a function of all the problem variables.
The heuristic policies accept people based on the claimed status, whereas the various classes of MDPs adapt their output actions based on the time and capacity remaining (shown in \cref{fig:mdp_policy}), resulting in a higher cumulative reward (shown in \cref{tab:results}).

\begin{figure*}
    \vspace{0.08in}
    \centering
    \subfloat[\label{fig:level1}Level I: The policy is provided with the claimed status (in black) and is unaware of the true status (shown in gray parentheses).]
    {\resizebox{\textwidth}{!}{\includegraphics{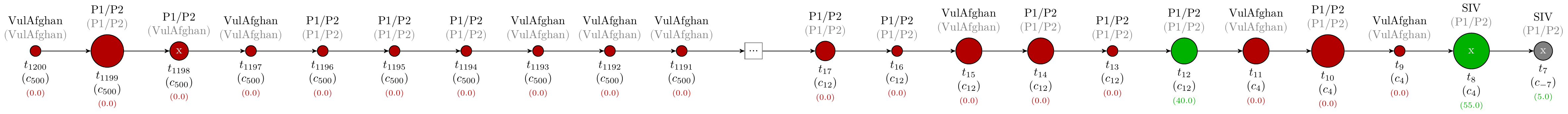}}}
    \hfill
    \subfloat[\label{fig:level2a}Level IIa: The uncertainty in the population distribution is updated over time and starts with the prior distribution shown above the left-most node.]
    {\resizebox{\textwidth}{!}{\includegraphics{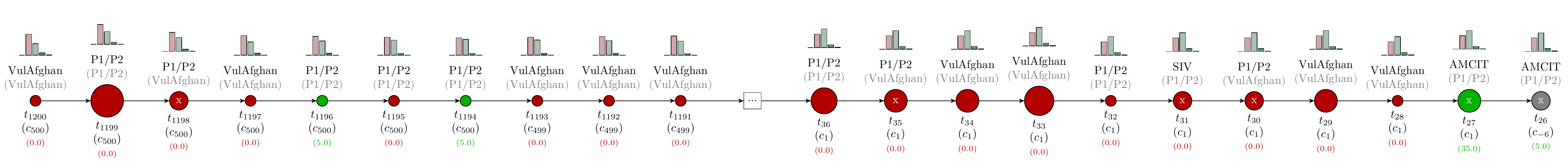}}}
    \hfill
    \subfloat[\label{fig:level2b}Level IIb: The uncertainty in the individual claimed status is used in the decision making (belief distributions over the statues are shown above each node).]
    {\resizebox{\textwidth}{!}{\includegraphics{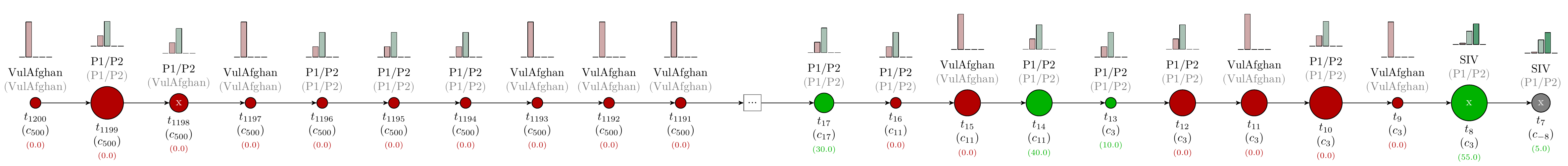}}}
    \hfill
    \subfloat[\label{fig:level3}Level III: Both uncertainties in the population distribution (top row of plots) and the individual claimed status (second row of plots) are used.]
    {\resizebox{\textwidth}{!}{\includegraphics{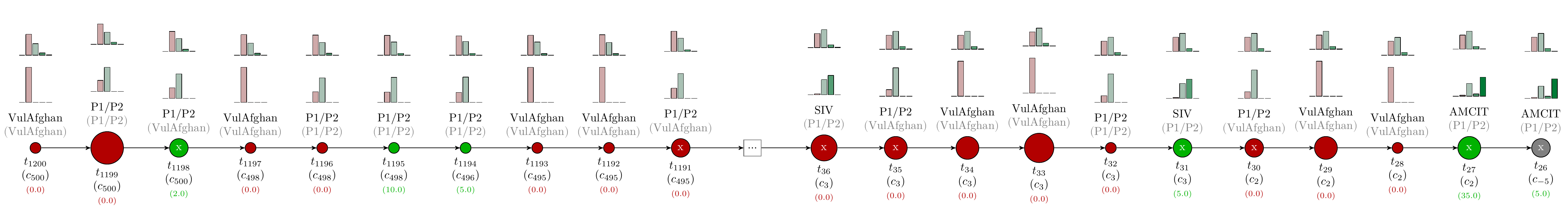}}}
    \caption{Decisions from a single simulation shown as a trajectory. Observed (claimed) priority status shown in black text with the true priority status shown in gray parentheses (hidden). A white ``x'' inside a node indicates a mismatch between the claimed and true priority status. The size of each node indicates the family size at the gate while the color indicates the \action{ACCEPT} or \action{REJECT} decision from the policy. The current time $t_i$, capacity $c_i$, and reward (in green/red) is shown below each node. The trajectory is split between the first ten time steps and the final ten time steps to illustrate the low-level behavior of each policy.}
    \label{fig:trajectories}
\end{figure*}

MDP-generated policies ultimately accept more people across claimed categories than the heuristic policies and end up with the highest cumulative reward shown in \cref{tab:results}.
We evaluated the performance of the policies on the 1000 simulated population sets and calculated the average number of people accepted by status category and reported the standard error.
\Cref{fig:trajectories} illustrates example trajectories of each policy on one simulated population set.
All policies were provided with the claimed status, but the reward is calculated with respect to the underlying true state.
The optimized policies generally have a higher total number of people accepted.
In heuristic policies, the total number of people accepted is lower than all the optimized polices.

\begin{figure}
    \centering
    \resizebox{0.99\linewidth}{!}{\includegraphics{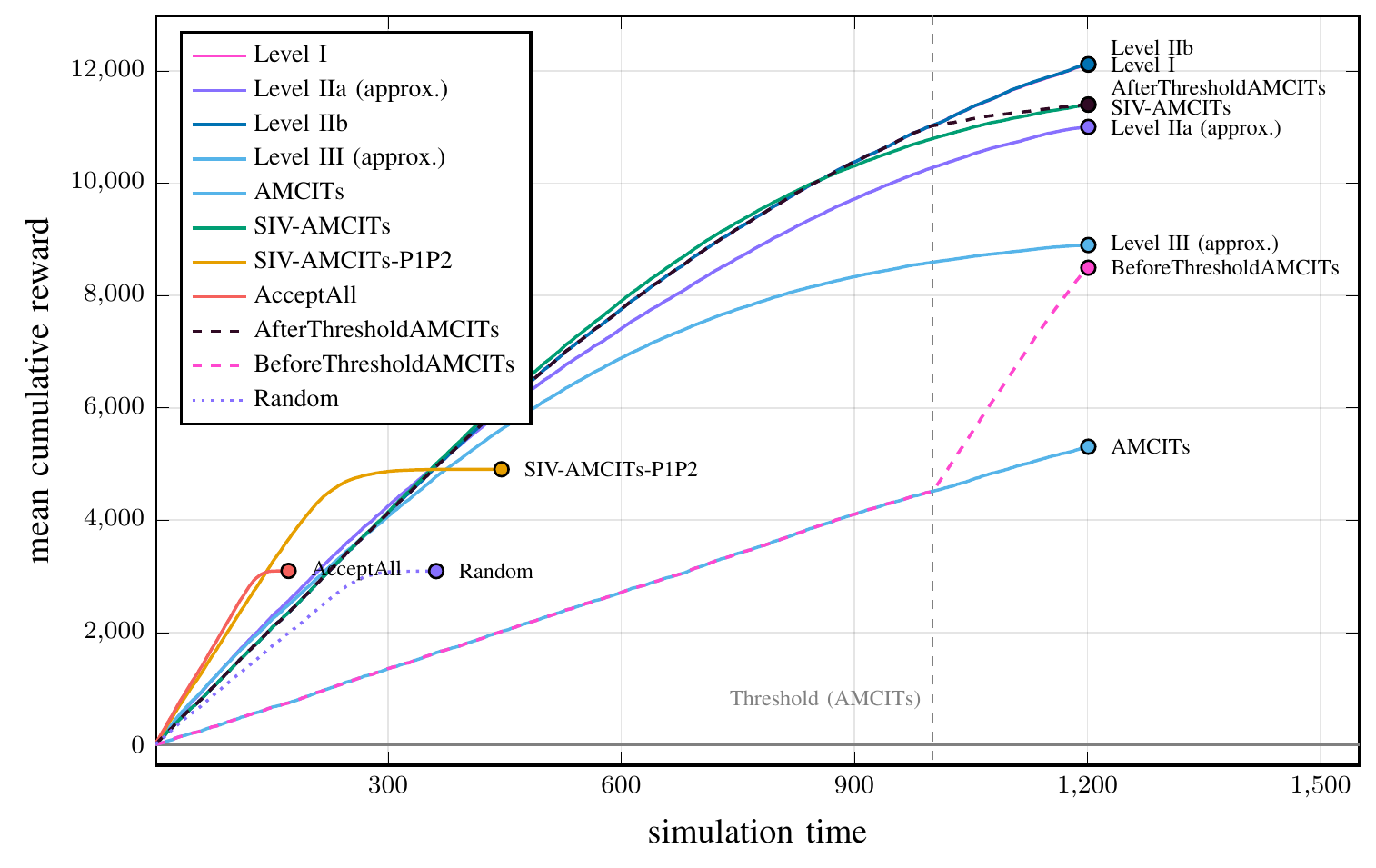}}
    \caption{Aggregate cumulative reward for each policy.}
    \label{fig:cumulative_reward}
\end{figure}

The compounding levels of uncertainty in the MDPs add complexity without improvement in policy performance.
We can see in \cref{fig:cumulative_reward} that level I (in which we assume we know the rough population distribution and priority status) and level IIb (in which we assume we know the rough population distribution but integrate our claims model to model uncertainty in priority status) result in the highest cumulative rewards and are nearly equal.
From this we can see that accounting for additional uncertainty in the individual status does not necessarily help the decision makers while involving subjective judgments on the claims model.
This suggests that trusting people's claims ultimately enhances the overall success of an evacuation, assuming one puts any value at all on categories of people who are not AMCITs or SIVs.
In addition to not increasing cumulative reward, models of claimed statuses are very subjective and use approximate POMDP solvers, which add approximation errors and runtime overhead.
 
Our experiments also help us see the limitations of rule-based policies such as accepting only AMCITs and SIVs. While accepting only AMCITs does result in the highest number of AMCITs accepted into the airport, the MDP-generated policy results in over twice as high a reward (shown in \cref{tab:results}). \Cref{fig:cumulative_reward} shows there are cases where early in the trajectory we get more reward (such as \textit{SIV-AMCITs-P1P2} or \textit{Accept All} policies) but these heuristic policies are myopic and underperform the more flexible MDP model over the full time horizon of the problem.


\subsection{Extracting heuristics}
\begin{figure}[h!]
  \resizebox{\linewidth}{!}{\includegraphics{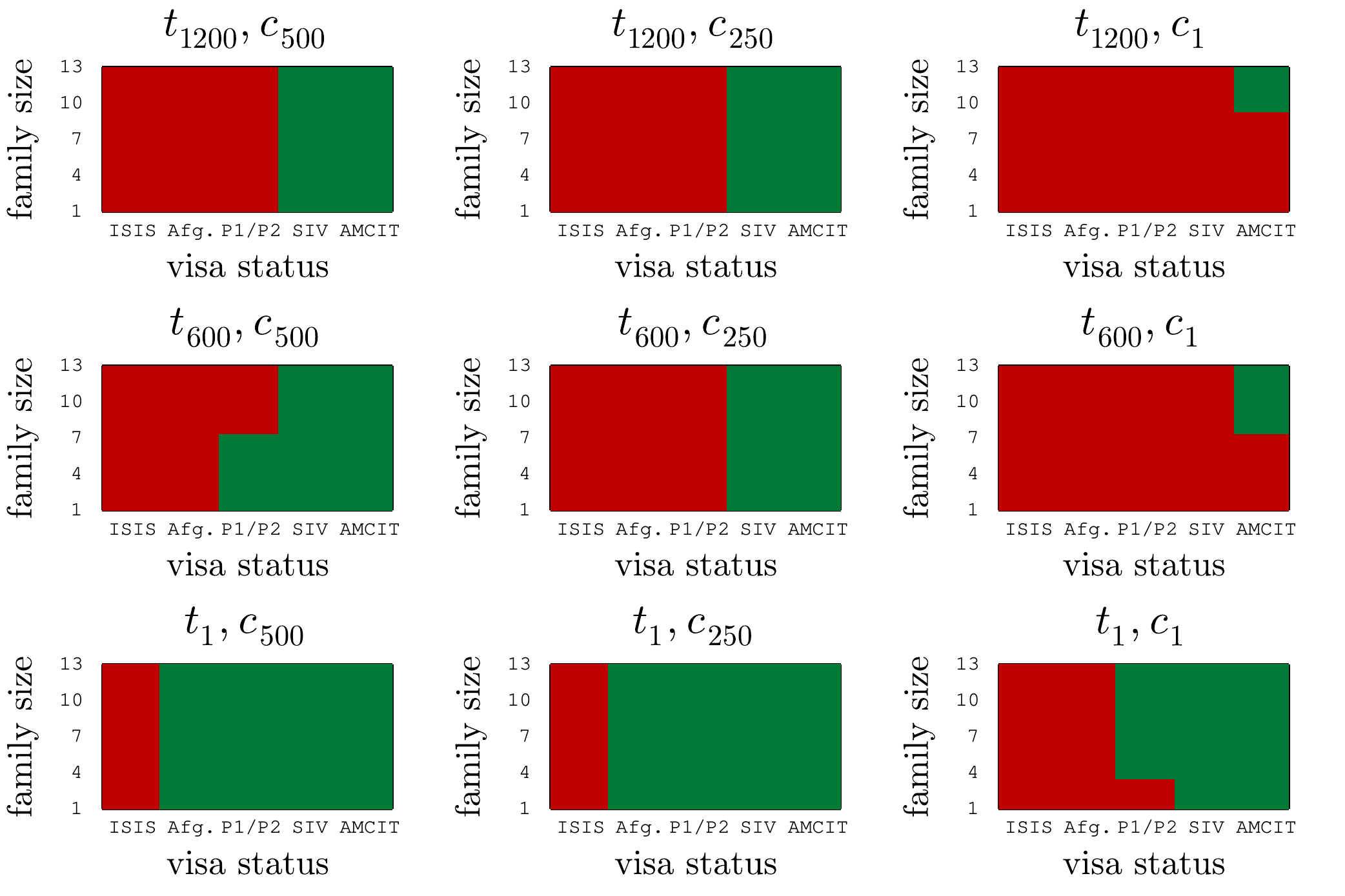}}
  \caption{Actions from the MDP policy when sweeping time and capacity. Note that green indicates \action{ACCEPT} and red indicates \action{REJECT}.}
  \label{fig:mdp_policy}
\end{figure}

Useful heuristics regarding how a policy can change with respect to time and capacity can be extracted from the optimized MDP policy to inform human decision makers.
Examining the MDP policy more closely, illustrated in \cref{fig:mdp_policy}, we can see the high-level behavior as time (rows) and capacity (columns) change.
When there is full time and capacity (top left plot), SIVs and AMCITs are the only categories accepted.
As time decreases but capacity remains high, the policy accepts larger family sizes of P1/P2 and vulnerable Afghans.
If time remains high and capacity is low, only AMCITs are prioritized.
If both time and capacity are low, the policy accepts SIVs, AMCITs, and those who claim P1/P2 status.
The policy is not intended to be used in an automated manner during an emergency, but it can be used as a learning tool for thinking about how decision making behavior could change with capacity and time.

\section{Discussion}
Sequential decision-making models could be used as a pre-crisis decision support system to engage military personnel and others involved in evacuations in reflections on policies for evacuation.
High-level officials from the Department of Defense have called for increased research into human-machine teaming and interaction and have called for integration of machine learning and deep learning-enabled applications into exercises, wargames, and tabletop exercises \cite{flournoy2020building}.
The policies developed through sequential decision-making frameworks can be used as a learning and discussion tool for policymakers in pre- and post-analysis of planning exercises.
For example, in this emergency setting, policymakers might assume integrating a claim model would result in a higher cumulative reward.
However, even if we include uncertainty in the individual's claimed status when calculating the reward for accepting that person, we achieve about the same reward as in the simpler policy.
There are of course additional security trade-offs inherent in trusting people's claimed statuses.
These trade-offs could be brought up in a pre-crisis debate, while including quantitative information on the ways conservative perspectives on claimed status could reduce the overall number of people saved. 

It is worth asking whether such a tool is actually practical for decision making or whether it (a) tells us strategies we already know or (b) is irrelevant given the chaos inherent in evacuation situations.
Even when we encode state uncertainty and treat the problem as a POMDP, there is still significant uncertainty in this multi-dimensional challenge.
We believe that an exercise like the one we describe in this paper could be useful \textit{ahead} of a crisis rather than during one.
The intelligence community and the military already model various scenarios ahead of conflicts. We offer this method as an additional framework that could be used in settings that involve sequential decision making while balancing multiple objectives.
These simulations would likely spark useful discussions and reflections about trade-offs, even if real-world policy choices would depend on shifting real-world circumstances.

Algorithmic frameworks should be developed and discussed in collaboration with ethicists and local experts.
This project highlights how difficult ethical decisions arise when values and priorities are assigned to human lives. Changing the reward assigned to a claimed status changes the number of people with that status who are accepted and rejected.
Likewise, leaders' decisions about prioritization of various categories of potential evacuees \textit{implicitly} suggest a reward function with values for each category of evacuee. For example, a policy that an evacuation will only evacuate American citizens can be optimal only if the reward assigned to all other categories of evacuees is small, zero, or negative. The use of algorithmic modeling in training decision makers may mitigate the harmful effects of cognitive biases \cite{sunstein2019algorithms} as they allow for new levels of transparency about difficult trade-offs implicit in human decisions \cite{kleinberg2018discrimination}.

We should be very cautious about fully replacing human decision makers with algorithms, given the many scenarios in which augmenting human judgement with algorithms can go wrong. For example, in 2020, Stanford Hospital used an algorithm to assign which employees would receive the COVID-19 vaccine first \cite{stanfordhospital}.
The algorithm rewarded older ages, which meant the algorithm did not equitably distribute vaccines to younger residents, even though they worked more closely with COVID-19 patients.
In high-stakes scenarios, algorithms are best used as decision-support or training tools, informed by diverse experts and local communities. Careful and critical study before and during the deployment of algorithmic decision support tools can help avoid unintended consequences. 

\section{Conclusion}
We proposed and demonstrated a principled way to answer the question of how to best evacuate a large population in a crisis situation, focusing on the 2021 Afghanistan evacuation.
We used different classes of Markov decision processes to capture compounding levels of uncertainty and compared their performance with eight heuristic policies.
We showed that the optimized MDP policy achieves greater performance than the heuristics.
Furthermore, complicating the model with increased uncertainty does not improve policy performance.
Our insights could be used to enrich pre-crisis policy discussions or in hindsight analysis, and possibly in human-AI joint-training exercises.
We open-sourced all tools and encourage researchers to build upon and strengthen our methods.

This preliminary work seeds a number of questions to be explored more thoroughly in future work. 
How should we handle evacuation cases with multiple gates or checkpoints?
Government officials present during the Afghanistan evacuation described challenges involving dangerous surges of people at particular gates.
The acceptance rate influenced the flow of people arriving at a particular gate, which would cause the gates to be 
closed. 
Another question regards how to classify priority status categories.
More granular status categories could also be included in future iterations of this work, such as Legal Permanent Resident (LPR) status or gender. Finally, we encourage others to explore the task of integrating such tools into the policymaking process.

\section*{Acknowledgment}
We thank Thomas Billingsley and Wren Elhai for their critical insights. We are also grateful to Patricia Wei and Lilian Chan for their support on early conceptualization of the project. 
\printbibliography

\end{document}